# Charge Conjugation and Parity Violations as a Signature for Black Hole Formation or Other New Physics in Hadron Collisions*

Michael J. Longo†

*Department of Physics, University of Michigan, Ann Arbor, MI 48109-1120 USA*

I point out that there have been essentially no tests of discrete symmetries, such as baryon number, charge conjugation (*C*), parity (*P*), strangeness, isospin, *etc.*, in high energy, high $p_T$ hadron collisions. If, for example, black hole formation (BHF) occurs, we might expect large violations of *C*, *P*, ... I propose new tests of *C* and *P* that can be adapted to a variety of new physics scenarios by selecting events with appropriate topologies. Large effects, such as ~10% longitudinal polarizations of outgoing particles, might be expected in events involving BHF. These tests may provide more sensitive searches for new physics at existing colliders.

Black hole formation in hadron collisions is a subject of great current interest. (See, for example, Refs. 1,2,3,4,5,6,7.) Speculation about catastrophic black hole formation at colliders has even appeared in the popular press. Though the possibility of such a spectacular signature as destroying the Earth has been disproved [8], more subtle signals of black hole formation in hadron collisions have proven to be elusive.

Most of the recent discussion of BHF has been in the context of the CERN Large Hadron Collider, though potentially observable rates could occur at the Fermilab Tevatron if the fundamental gravity scale is below 1.4 TeV. [3] Proposed signatures for BHF range from lepton production at large transverse momenta[2] to the presence[4] (or absence[5]) of multijet events with large transverse momenta. The formation of a quark-gluon plasma decaying into thousands of photons has also been suggested.[7] Originally, it was thought that BHF events would show large missing energy due to graviton radiation[1], while more recently Emparan, Horowitz, and Myers[6] have shown that the radiation is mainly on the brane. No experimental evidence of BHF has yet been found, but, particularly with these somewhat contradictory scenarios, it is important to explore all possible signatures.

---

It has long been believed that only quantum numbers and symmetries that are associated with long-range fields are preserved in black hole formation.[8] Thus, charge, angular momentum, and mass are presumed to be conserved, while baryon number and lepton number, as well as charge conjugation, parity, *CPT*, isospin, strangeness and other quantum numbers that are believed to be conserved only in hadronic interactions, can be expected to show large violations if BHF and subsequent decay by Hawking radiation[8] occurs in hadron collisions. Here I discuss how such signatures of black hole formation or other new physics might manifest themselves and how they might best be observed (or perhaps might already have been observed).

Black hole formation or other new physics is likely to occur in the most violent quark-quark collisions. Thus, the most promising venue for finding BHF before the CERN LHC is the Fermilab Collider. Because hard quark-quark collisions are generally associated with the production of particles at large transverse momenta or with high multiplicities, if BHF occurs we might expect large violations of baryon number, charge conjugation, strangeness, isospin, *etc.* in such processes. While violations of baryon or lepton number conservation would be a very dramatic signature, as a practical matter all the existing collider detectors have large uninstrumented regions where most of the particles from interactions are lost. This limitation, as well as the lack of particle species identification except for low momentum particles, makes direct tests of baryon/lepton number conservation, strangeness, *etc.*, impractical. However, the unique properties of BHF and subsequent decay could lead to correlations in events that would be very hard to explain by any other mechanism. For example, striking events with 2 (or more) like-sign *W*'s or high $p_T$ muons could occur from black hole decays with a probability comparable to that for opposite sign.

One promising method to search for BHF appears to be testing charge conjugation (*C*) in hadronic interactions at large momentum transfers. Ironically, there have been very few tests of *C* in strong interactions generally,[9] and none at large momentum transfers or collider energies. This is surprising as a $\bar{p}$-$p$ collider with charge conjugate particles of equal momentum in the initial state is an ideal laboratory for testing *C* invariance.

An immediate prediction of *C* invariance is that the spectrum of any charged particle in the forward $\bar{p}$ hemisphere should be identical to that of the charge conjugate particle in the forward $p$ hemisphere. This can be tested in a collider detector with a magnetic field, even without particle identification. The only relevant data appear to be from the CDF detector. These unpublished data from the thesis of A. Byon[10] are reproduced in Fig. 1. In this figure the pseudorapidity $\eta \equiv -\log_e(\tan\frac{\theta}{2})$, so that $\eta = 0$ corresponds to 90° with respect to either beam. Thus the expectation of charge conjugation is that the positive and negative distributions should be mirror symmetric about $\eta = 0$. Unfortunately, the numerical data are no longer available, but a careful evaluation of the graphed data shows no sign of any *C*-violating asymmetry. However, it should be noted that these data are at low transverse momentum, $\bar{p}_t \approx 1$ GeV/c, with no special event selection, so are not a good test for new physics. A careful test should be done with high $p_T$ particles and would require corrections for apparent $C$ violations due to weak decays, such as $\Lambda \to p + \pi^-$, which generally show large violations.

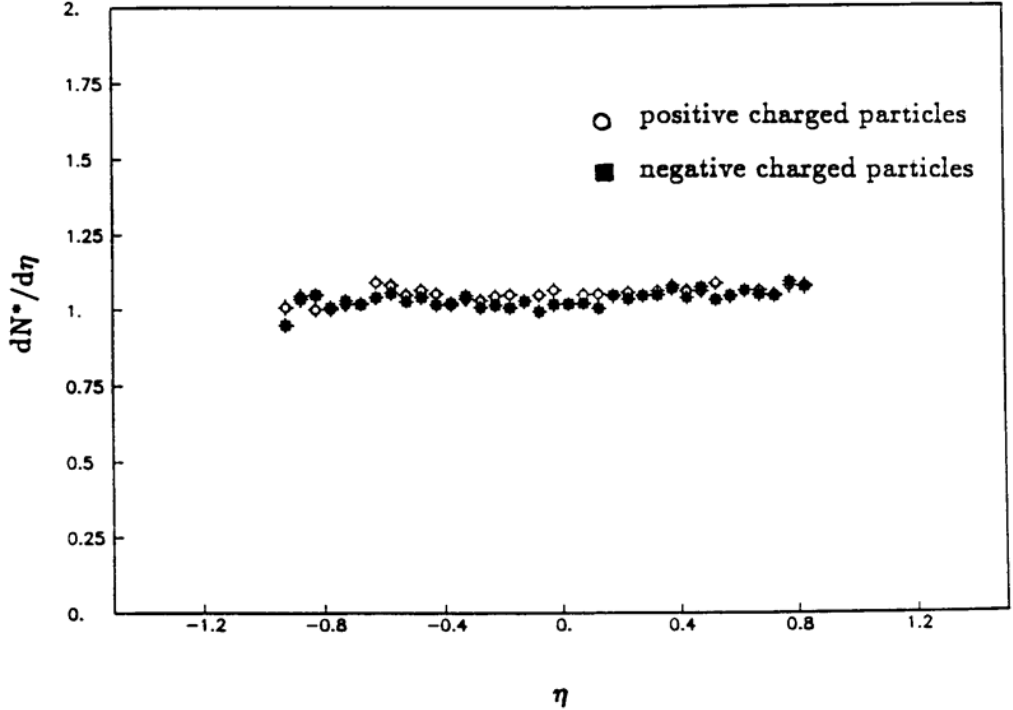

**FIGURE 1** Distribution of positive and negative particles from central collisions in the CDF detector from the thesis of A. Byon[13]. $\eta \equiv -\log_e(\tan\frac{\theta}{2})$. Charge conjugation requires that they be mirror symmetric about $\eta = 0$.

Large parity violations might also occur due to BHF. This could be tested by searching for parity-violating spin components of hyperons such as $\sigma \cdot p$ where $p$ is the vector momentum of the hyperon. The large asymmetry in $\Lambda^0$ hyperon decays makes this a particularly attractive possibility for a sensitive search for BHF. In this search one can choose whatever sample of events that are most likely to be associated with BHF, such as high multiplicity, high (or low) transverse momentum, high (or low) missing energy, *etc.* Another possibility is to look for parity-violating correlations such as a term $p_1 \cdot (p_2 \times p_3)$, where $p_1, p_2, p_3...$ are any combination of beam particle, lepton, photon, $W^\pm$, $Z^0$, high $p_T$ particle, or jet momenta.[11] Again, a variety of event types and correlations can be looked at to investigate a variety of new physics scenarios.

While it is not possible to estimate the size of $C$ or $P$ violating effects or baryon and lepton number violations in BHF without a real theory of quantum gravity, it can be argued that, in principle, such effects can be quite large, of order unity. This is easiest to see if we consider the example of the time-inverse process of a neutrino about to fall through the event horizon of a black hole, as illustrated in Fig. 2. It is clear that the situation in (b) is not the charge conjugate of that in (a), and that lepton number is violated. Thus it is plausible to expect that whatever mechanism produces the charge conjugation and lepton number violations in black hole formation will also lead to large violations of lepton number, baryon number, and $C$ in the inverse process in which particles are produced in the decay of a black hole. In a decay with few particles in the final state, these effects should be of order unity. In an event with $N$ particles in the final state, we might naively expect effects $\sim 1/N$. The expected multiplicity from BHF, even at LHC energies, is $\sim 6$ in some scenarios[2, 6], so these effects are still sizeable.

By similar arguments we can conclude that $P$, $T$, $CP$, and $CPT$ [at least in the sense they are normally used in particle physics] are also likely to show large violations in decays of black holes produced in high-energy collisions. The mirror symmetry of charged particle production discussed above as a test of $C$ is also required by $CP$.

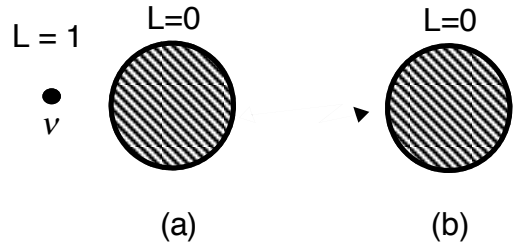

**FIGURE 2.** (a) A neutrino about to enter the event horizon of a black hole; (b) The system formed just after is not the charge conjugate of that in (a). Thus, charge conjugation is violated maximally when the neutrino falls into the black hole, and it is reasonable to expect that quantum gravity will show comparably large violations when particles are produced in black hole decay.

Though violations of causality and $CPT$ might be considered the holy grail of BHF, it is not obvious how to test these. It is not possible to test for isospin violations due to BHF directly. However, if BHF does occur, there is no constraint on isospin in the final state, so that large fluctuations in the ratio of charged to neutral pions might be expected. Such fluctuations have been observed in cosmic ray events ("Centauro" and "anti-Centauro" events with few $\pi^0$ and mostly $\pi^0$, resp.).[12] These could be a signal for BHF, though many other interpretations have been suggested.

Tests of $C$ and $P$ at large $p_T$ from electron-positron collisions are also of interest. The apparent absence of electric dipole moments of the electron and neutron[12] are excellent tests of $P$ in electromagnetic interactions. Tests of $C$ in electromagnetic interactions have been made[12] by searching for $C$-violating decays of low-mass particles such as $\eta^0 \rightarrow \pi^0 + \mu^+ + \mu^-$, but no tests of $C$ or $P$ have been made at large transverse energies.

I conclude that violation of discrete symmetries, such as *C* and *P*, may be the best "smoking gun" signature for new physics in hadronic interactions. Relatively straightforward experimental tests are possible, very few of which have been done. In any case, these symmetries have never been tested at large transverse momenta, and large violations could occur from black hole formation or other "new physics". Causality and *CPT* violation could also occur, though it is not clear how these would manifest themselves or be tested for.

If BHF can be identified and their properties studied through their violations of symmetries and conservation laws, it will provide a long sought connection between quantum mechanics and general relativity. Copious production of black holes is predicted at the LHC; they may already be produced at observable rates at the Fermilab Tevatron. All possible tests should be done on Fermilab data while we eagerly await results from the LHC.